\documentclass[aps,reprint,prd,showpacs,nofootinbib,twocolumn]{revtex4}

\usepackage{amsmath,amssymb}
\usepackage{graphicx,subfigure}
\usepackage{color,multirow}
\usepackage[colorlinks,linkcolor=magenta,anchorcolor=cyan,citecolor=blue,plainpages=false]{hyperref}

\hypersetup{colorlinks=true,
    breaklinks=true,
    pdfstartview=Fit,
    linkcolor=blue,
    citecolor=blue,
    urlcolor=blue}

\def\be{\begin{equation}}
    \def\ee{\end{equation}}
\def\ba{\begin{eqnarray}}
    \def\ea{\end{eqnarray}}

\begin{document}

\title{Can recent DESI BAO measurements accommodate a negative cosmological constant?}

    \author{Hao Wang$^{1,2} $\footnote{\href{wanghao187@mails.ucas.ac.cn}{wanghao187@mails.ucas.ac.cn}}}
    \author{Ze-Yu Peng$^{3} $ \footnote{\href{pengzeyv@mail.ustc.edu.cn}{pengzeyv@mail.ustc.edu.cn}}}
    \author{Yun-Song Piao$^{1,2,3,4} $ \footnote{\href{yspiao@ucas.ac.cn}{yspiao@ucas.ac.cn}}}

    \affiliation{$^1$ School of Fundamental Physics and Mathematical
        Sciences, Hangzhou Institute for Advanced Study, UCAS, Hangzhou
        310024, China}

    \affiliation{$^2$ School of Physics Sciences, University of
        Chinese Academy of Sciences, Beijing 100049, China}

    \affiliation{$^3$ International Center for Theoretical Physics
        Asia-Pacific, Beijing/Hangzhou, China}

    \affiliation{$^4$ Institute of Theoretical Physics, Chinese
        Academy of Sciences, P.O. Box 2735, Beijing 100190, China}

    \begin{abstract}
Anti-de Sitter vacuum, which correspond to a negative cosmological
constant (CC), is theoretically important and well-motivated.
However, whether it exists in reality or not has always been a
controversial issue. In this paper, we perform the search for the
negative CC using recent Dark Energy Spectroscopic Instrument
(DESI) baryon acoustic oscillation measurements combined with
Planck cosmic microwave background and Pantheon Plus supernova
data. Though we did not find the evidence for negative CC, we
observed the indication for it, the negative CC is preferred at
$>68\%$ significance level, while such a negative CC can make the
state equation of evolving dark energy component (coexisting with
negative CC) $w\geqslant -1$. Our work highlights the potential of
upcoming cosmological surveys to search for the negative CC.


    \end{abstract}

    \maketitle

\section{Introduction}

The concordance $\Lambda$CDM model is the simplest description of
our universe, which can naturally explain most of cosmological and
astrophysical observations. In corresponding model, $\Lambda$,
also called the dark energy (DE), is responsible for the
accelerated expansion of current universe. In the past decades,
identifying the nature of DE has been still an important
challenge, see
e.g.\cite{Carroll:2000fy,Frieman:2008sn,Nojiri:2017ncd,Huterer:2017buf}
for reviews. It is usually thought that DE is a positive
cosmological constant (CC),
e.g.\cite{Weinberg:1988cp,Padmanabhan:2002ji}. However, recently
DESI collaboration \cite{DESI:2024lzq,DESI:2024mwx,DESI:2024uvr}
using their first year baryon acoustic oscillation (BAO) data,
combined with Planck cosmic microwave background (CMB) data, has
found $w_0=-0.45\pm0.27$ and $w_a=-1.79\pm0.79$ for the
parameterised equation of state of DE
\cite{Chevallier:2000qy,Linder:2002et}
\begin{equation}
    w(z)=w_0+w_a\frac{z}{1+z},
    \label{wxz}
\end{equation}
while when Pantheon data is considered, the constraints on $w_0$
and $w_a$ becomes tighter, $w_0=0.827\pm0.063$ and
$w_a=-0.75\pm0.27$, the CC ($w_0=-1$ and $w_a=0$) has been not
preferred at $>2\sigma$ C.L. see also
\cite{DESI:2024aqx,DESI:2024kob}. The relevant issues have been
also intensively investigated recently,
e.g.\cite{Yang:2024kdo,Wang:2024dka,Wang:2024pui,Yin:2024hba,Luongo:2024fww,Cortes:2024lgw,Carloni:2024zpl,Wang:2024hks,Wang:2024rjd,Colgain:2024xqj,Giare:2024smz,Gomez-Valent:2024tdb,Escamilla-Rivera:2024sae,Park:2024jns,Shlivko:2024llw,Dinda:2024kjf,Seto:2024cgo,Bhattacharya:2024hep},
see also \cite{Wolf:2023uno}.

It is well-known that anti-de Sitter (AdS) vacuum, which
correspond to a negative CC, is theoretically important and
well-motivated. Though it is possible that dS vacua exist
\cite{Kachru:2003aw,Kallosh:2004yh}, the construction of such
vacua in string landscape is not straightforward, notably e.g.
recent swampland conjecture \cite{Ooguri:2006in,Obied:2018sgi},
see \cite{Palti:2019pca,Grana:2021zvf} for recent reviews and also
\cite{Kallosh:2019axr}. However, AdS vacua exist naturally in
string landscape.


Recently, the impacts of AdS vacua on our observable universe have
been investigated, in particular the resolution of the Hubble
tension might be related to the AdS vacuum around the
recombination
\cite{Ye:2020btb,Ye:2020oix,Jiang:2021bab,Ye:2021iwa,Wang:2022jpo},
or the switch-sign CC
\cite{Akarsu:2019hmw,Akarsu:2021fol,Akarsu:2022typ,Akarsu:2023mfb,Paraskevas:2024ytz},
see also \cite{Anchordoqui:2023woo}. The inflation in AdS
landscape and its implication for primordial perturbations have
been explored in
e.g.\cite{Li:2019ipk,Ye:2022efx,Lin:2022ygd,Piao:2005ag,Cai:2017pga}.
Though a negative CC is unable to drive the acceleration of our
universe, it can coexist with one evolving positive DE component,
and the corresponding effects have been surveyed in light of
cosmic microwave background (CMB) and pre-DESI BAO dataset
\cite{Dutta:2018vmq,Visinelli:2019qqu,Ruchika:2020avj,Calderon:2020hoc,Sen:2021wld,Malekjani:2023ple,Gomez-Valent:2023uof},
as well as recent James Webb Space Telescope (JWST) observations
\cite{Adil:2023ara,Menci:2024rbq}.

Though the existence of AdS vacua could profoundly affect our
understanding about the nature of DE and quantum gravity, whether
it exists in reality or not has always been a controversial issue.
In this paper, we perform the first search for the negative CC
using recent DESI BAO measurements combined with Planck CMB and
Pantheon Plus supernova data. Though we did not find the evidence
for negative CC, we observed the indication for it, the negative
CC is preferred at $>68\%$ significance level, see Fig.1.



\begin{figure}
\includegraphics[width=0.7\columnwidth]{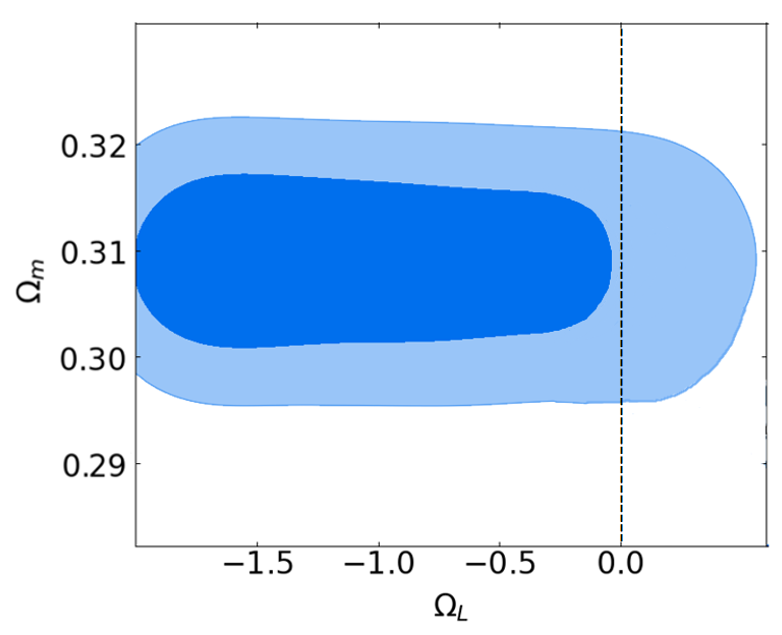}
\caption{\label{OmegaL} The 1$\sigma$ and $2\sigma$ contours of
$\Omega_L-\Omega_m$ with the Planck18+DESI+Pantheon Plus dataset,
where $\Omega_m$ is the current fraction of matter density and
$\Omega_L$ is that of the CC.  }
\end{figure}

\section{Data and Methodology}

To perform the search for the
negative CC using recent observations, we adopt the following
observational datasets:
\begin{itemize}
    \item [$\bullet$] \textbf{CMB} Planck 2018 low-l and high-l TT,
    TE, EE spectra, and reconstructed CMB
    lensing spectrum \cite{Planck:2018vyg,Planck:2019nip,Planck:2018lbu}.
    \item [$\bullet$] \textbf{DESI BAO}
    The measurements of DESI for the comoving distance $D_M/r_d$
and $D_H/r_d$ \cite{DESI:2024mwx,DESI:2024lzq,DESI:2024uvr},
    \begin{equation}
        D_M(z)\equiv\int_{0}^{z}{c dz'\over H(z')},\quad D_H(z)\equiv {c\over
        H(z)},
    \end{equation}
where $r_d=\int_{z_d}^{\infty}{c_s(z)\over H(z)}$ is the sound
horizon, $z_d\simeq1060$ is the redshift at the baryon drag epoch
and $c_s$ is the speed of sound, as well as the angle-averaged
quantity $D_V/r_d$ where
$D_V(z)\equiv\left(zD_M(z)^2D_H(z)\right)^{1/3}$.

    \begin{table}[htbp]
        \centering
        \begin{tabular}{c|c|cc|c}
            tracer&$z_\mathrm{eff}$&$D_M/r_d$&$D_H/r_d$&$D_V/r_d$\\
            \hline
            BGS&0.30&-&-&$7.93\pm0.15$\\
            LRG&0.51&$13.62\pm0.25$&$20.98\pm0.61$&-\\
            LRG&0.71&$16.85\pm0.32$&$20.08\pm0.60$&-\\
            LRG+ELG&0.93&$21.71\pm0.28$&$17.88\pm0.35$&-\\
            ELG&1.32&$27.79\pm0.69$&$13.82\pm0.42$&-\\
            QSO&1.49&-&-&$26.07\pm0.67$\\
            Lya QSO&2.33&$39.71\pm0.94$&$8.52\pm0.17$&-\\
        \end{tabular}
    \caption{\label{DESI}Statistics for the DESI samples of the DESI DR1 BAO measurements used in this
            paper \cite{DESI:2024mwx}.}
    \end{table}
    \item [$\bullet$] \textbf{Pantheon+} We use Pantheon+ consisting
    of 1701 light curves of 1550 spectroscopically confirmed type Ia
    SN coming from 18 different surveys \cite{Scolnic:2021amr}.
\end{itemize}

Here, we assumed that one CC and one evolving positive DE
component coexist, called the $w_0w_a$CDM+CC model for simplicity,
where the state equation $w_x(z)$ of positive-energy component of
DE follows (\ref{wxz}). They are related by
$\Omega_m+\Omega_r+\Omega_x+\Omega_\Lambda=1$, where the current
fraction of the matter density $\Omega_m$, the radiation
$\Omega_r$, the evolving positive-energy component of DE
$\Omega_x$ and the CC $\Omega_L$.

The parameter space of the $w_0w_a$CDM+CC model adapted is
\{$100\omega_b$, $\omega_{cdm}$, $H_0$, $n_s$, $\ln10^{10}A_s$,
$\tau_{reio}$, $w_0$, $w_a$, $\Omega_x$\}. There is only an
additional parameter $\Omega_x$, compared with the $w_0w_a$CDM
model, while that of the CC component is related to $\Omega_x$ by
$\Omega_\Lambda=1-\Omega_m-\Omega_r-\Omega_x$. We modified the
MontePython-3.6 sampler \cite{Audren:2012wb,Brinckmann:2018cvx}
and CLASS codes \cite{Lesgourgues:2011re,Blas:2011rf} to perform
our MCMC analysis. In Table.\ref{prior} we list the flat priors
used. We adopt a Gelman-Rubin convergence criterion with a
threshold $R-1<0.01$.

\begin{table}[htbp]
    \centering
    \begin{tabular}{cc}
        \hline
        Parameter&Prior\\
        \hline
        $100\omega_b$&[None, None]\\
        $\omega_{cdm}$&[None, None]\\
        $H_0$&[65, 80]\\
        $\ln10^{10}A_s$&[None, None]\\
        $n_s$&[None,None]\\
        $\tau_{reio}$&[0.004, None]\\
        \hline
        $w_\mathrm{0,eff}$&[-2, 0.34]\\
        $w_\mathrm{a,eff}$&[-3,2]\\
        $\Omega_L$&[-2,0.6]\\
        \hline
    \end{tabular}
    \caption{\label{prior} The priors of parameters we adapt in MCMC analysis. $w_\mathrm{0,eff}$ and $w_\mathrm{a,eff}$ are defined in (\ref{w0eff}) and (\ref{waeff}).}
\end{table}

\section{Results}

In Table.\ref{DE} and \ref{chi2}, we present our
MCMC results and the corresponding $\chi^2$.
As in the $w_0w_a$CDM model, for the $w_0w_a$ component of DE,
$w_0=-1$ and $w_a=0$ is still ruled out beyond $\gtrsim 2\sigma$
C.L.. However, a negative CC (the bestfit $\Omega_\Lambda\sim
-0.3$) is preferred at $>1\sigma$ C.L. by data (the fit is
improved with $\Delta\chi^2\sim-1.1$).

    \begin{table}[htbp]
    \centering
    \begin{tabular}{c|c|c}
        \hline  & $w_0w_a$CDM & $w_0w_a$CDM\\
        Parameters& &+CC\\
        \hline
        $100\omega_b$&2.219(2.207)$\pm$0.014&2.234(2.234)$\pm$0.015\\
        $\omega_{cdm}$&0.119(0.120)$\pm$0.001&0.120(0.119)$\pm$0.001\\
        $H_0$&67.86(67.99)$\pm$0.73&68.32(67.91)$\pm$1.02\\
        $\ln10^{10}A_s$&3.036(3.026)$\pm$0.014&3.038(3.034)$\pm$0.017\\
        $n_s$&0.964(0.961)$\pm$0.004&0.968(0.967)$\pm$0.004\\
        $\tau_{reio}$&0.052(0.048)$\pm$0.007&0.053(0.052)$\pm$0.007\\
        \hline
        $w_0$&-0.844(-0.886)$^{+0.073}_{-0.068}$&-0.857(-0.861)$^{+0.033}_{-0.107}$\\
        $w_a$&-0.670(-0.546)$^{+0.298}_{-0.311}$&-0.647(-0.609)$^{+0.658}_{-0.217}$\\
        $\Omega_L$&0&-0.879(-0.316)$^{+0.639}_{-1.012}$\\
        \hline
        $\Omega_m$&0.3087(0.3077)$\pm$0.0069&0.3025(0.3091)$\pm$0.0098\\
        $S_8$&0.833(0.832)$\pm$0.012&0.834(0.827)$\pm$0.011\\
        \hline
    \end{tabular}
    \caption{\label{DE} Mean (bestfit) values and 1$\sigma$ regions
of the parameters of $w_0w_a$CDM+CC model. The dataset used is the
Planck18+DESI+Pantheon Plus
        dataset.}
\end{table}

    \begin{table}[htbp]
    \centering
    \begin{tabular}{c|c|c}
        \hline  & $w_0w_a$CDM & $w_0w_a$CDM\\
        Datasets& &+CC\\
        \hline
        Planck high-l TTTEEE&2341.65&2341.20\\
        Planck low-l EE&396.10&395.81\\
        Planck low-l TT&22.55&22.54\\
        Planck lensing&8.77&8.79\\
        DESI&14.93&14.27\\
        Pantheon+&1409.66&1409.78\\
        \hline
        Total $\chi2$ &4193.76&4192.64\\
        \hline
    \end{tabular}
\caption{\label{chi2} $\chi^2$ of $w_0w_a$CDM and $w_0w_a$CDM+CC
model.}
\end{table}

\begin{figure}
    \includegraphics[width=0.85\columnwidth]{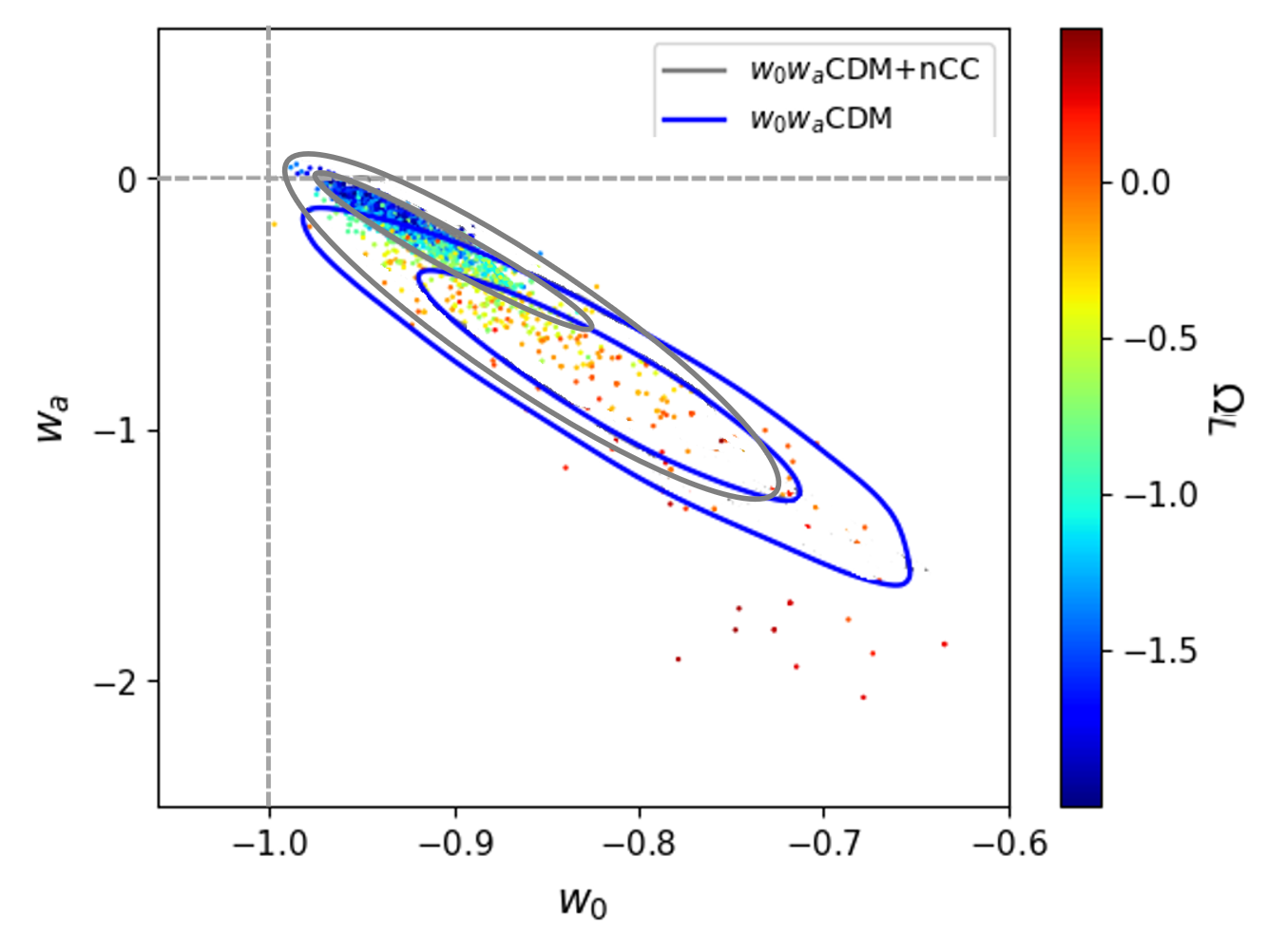}
\caption{\label{ws}1$\sigma$ and 2$\sigma$ contours of $w_0-w_a$
for the $w_0w_a$CDM and $w_0w_a$CDM+CC models, respectively, and
scattered plot for the $w_0w_a$CDM+CC model with respect to
$\Omega_\Lambda$.}
\end{figure}

In Fig.\ref{ws}, we find that compare to the $w_0w_a$CDM model,
the $2\sigma$ contour of $w_0-w_a$ in the $w_0w_a$CDM+CC models
shifts up as a whole, so that $w_a=0$ can be $1\sigma$ consistent
with the observations, which is an unexpected result. This can be
explained as follows. In our case, the low-redshift evolution of
background is co-controlled by the evolving DE component
$\Omega_x$ with the state equation $w_x$ and CC
$\Omega_{\Lambda}$, beside the matter, it is interesting to
consider an effective state equation of $w0w_a+$CC DE
\cite{Adil:2023ara}\footnote{As discussed in
Ref.\cite{Ozulker:2022slu,Adil:2023ara}, for initially $w_x(z)<-1$
and negative $\Omega_\Lambda$, $w_\mathrm{eff}(z)$ will diverge at
certain high redshift. Thus here (\ref{we}) only serves as an
effective description of state equation at low redshift. }
\begin{equation}
w_\mathrm{eff}(z)=\frac{\Omega_x(1+z)^{2+3w_0+3w_a}[w_0+(w_0+w_a)z]e^{-\frac{3w_az}{1+z}}-\Omega_{\Lambda}}{\Omega_x(1+z)^{3(1+w_0+w_a)}e^{-\frac{3w_az}{1+z}}+\Omega_{\Lambda}},\label{we}
\end{equation}
instead of $w_x$. Analogous to the parameterization (\ref{wxz}),
we have
\begin{equation}
    w_\mathrm{0,eff}\equiv w_\mathrm{eff}(z)|_{z=0}=\frac{\Omega_xw_0-\Omega_\Lambda}{\Omega_x+\Omega_\Lambda},\label{w0eff}
\end{equation}
and
\begin{equation}
    w_\mathrm{a,eff}\equiv \frac{dw_\mathrm{eff}(z)}{dz}|_{z=0}=\frac{\Omega_x[w_a\Omega_x+3(w_0+1)^2\Omega_\Lambda]}{(\Omega_x+\Omega_\Lambda)^2}. \label{waeff}
\end{equation}
It can be observed that for $\Omega_{\Lambda}<0$, we will have
$w_\mathrm{0,eff}>w_0$ and $w_\mathrm{a,eff}<w_a$ (noting
$\Omega_x+\Omega_\Lambda=1-\Omega_m-\Omega_r\sim 0.7$). Thus if
$w_\mathrm{0,eff}>-1$ and $w_\mathrm{a,eff}<0$ supported by
observations, $w_0$ and $w_a$ must be closer to $w_0=-1$ and
$w_a=0$, which is just the result showed in Fig.\ref{ws}.

The reason that an moderate negative CC with the bestfit
$\Omega_\Lambda\sim-0.3$ is preferred by Planck18+DESI+Pantheon
Plus dataset is that it suggests $w_\mathrm{0,eff}\sim-0.8$ and
$w_\mathrm{a,eff}\sim-0.7$ at low redshift, as reported by the
DESI collaboration for the $w0w_a$CDM model. However, the
inclusion of a negative CC in a positive $w_0w_a$DE component can
make $w_0>-1$ and $w_a\sim 0$ (so we can always have $w\geqslant
-1$) $1\sigma$ consistent. The negative CC component with
$|\Omega_\Lambda|\sim O(1)$ has been also considered in e.g.
Refs.\cite{Dutta:2018vmq,Visinelli:2019qqu,Sen:2021wld,Adil:2023ara}.

\begin{figure*}
    \includegraphics[width=1.6\columnwidth]{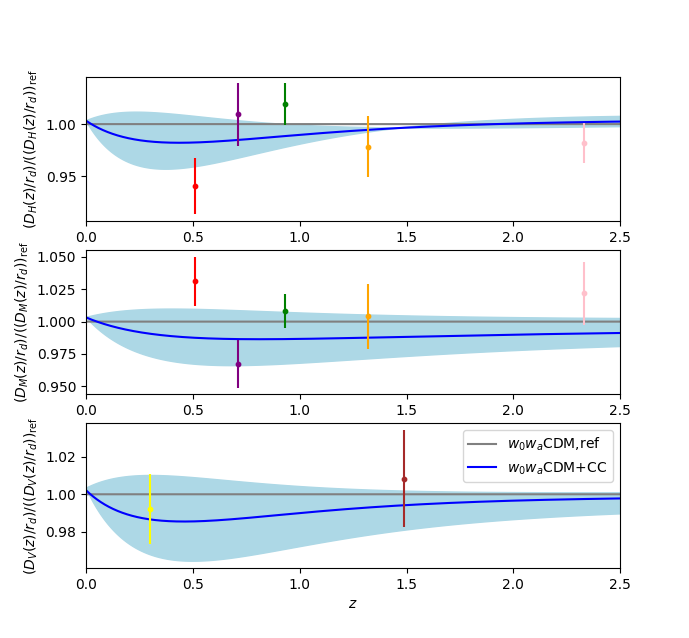}
\caption{\label{DH}Residuals of $D_{H}(z)/r_d$, $D_{M}(z)/r_d$ and
$D_{V}(z)/r_d$ for the bestfit values (blue curves) of the
$w_0w_a$CDM+CC model with respect to the $w_0w_a$CDM model. The
blue shadows are the corresponding $1\sigma$ regions. }
\end{figure*}

In Fig.\ref{DH}, we show the evolutions of $D_H(z)/r_d$,
$D_M(z)/r_d$ and $D_V(z)/r_d$ for our bestfit values, in which
$\Omega_\Lambda=-0.316$. It is found that a negative CC could
slightly suppress $D_H(z)/r_d$, $D_M(z)/r_d$ and $D_V(z)/r_d$ at
low redshift, and notably the negative CC can bring a slightly
better fitting at the effective redshift $z_\mathrm{eff}=0.51$
(see e.g.the comments in
Refs.\cite{Wang:2024pui,Wang:2024rjd,Colgain:2024xqj,Carloni:2024zpl}
on DESI data at $z_\mathrm{eff}=0.51$) and $z_\mathrm{eff}=0.71$,
which might be responsible to the improvement of
$\Delta\chi^2_\mathrm{DESI}\sim-0.6$, see Table.\ref{chi2}. This
result can be explained as follows. According to the
parameterization (\ref{wxz}), for a negative $\Omega_\Lambda$, we
have \ba {H^2(z)\over
H^2_0}&=&\Omega_m(1+z)^3-\left|\Omega_\Lambda\right|\nonumber\\
& &
+(1-\Omega_m+\left|\Omega_\Lambda\right|)(1+z)^{3(1+w_0+w_a)}e^\frac{-3w_az}{1+z}.\ea
Thus
$\rho_x(z)=\rho_0(1-\Omega_m+\left|\Omega_\Lambda\right|)(1+z)^{3(1+w_0+w_a)}e^\frac{-3w_az}{1+z}$
(so $H(z)$) must be larger than that in the $w_0w_a$CDM model
without negative CC, or else $\rho_x(z)+\rho_{\Lambda}$ at low
redshift will lower too fast to fit the observation. In certain
sense the concavities of $D_H(z)/r_d$, $D_M(z)/r_d$ and
$D_V(z)/r_d$ at low redshift might be an observable imprint of
negative CC (AdS vacuum), the upcoming more survey data would rule
out it.

\section{Discussion}

In this paper, with recent DESI BAO data we found the indication
for the negative CC, which is preferred at $>68\%$ C.L.. The
quintessence has been ruled out at $\gtrsim 2\sigma$ C.L. in
recent DESI paper \cite{DESI:2024mwx}, however, an unexpected
effect of our result is that the inclusion of a negative CC will
make $w_0$ and $w_a$ for the evolving DE component (coexisting
with negative CC) closer to $w_0=-1$ and $w_a=0$, in particular,
$w_0+w_a\geqslant -1$ (i.e. a quintessence-like component) can be
$1\sigma$ consistent.





In our phenomenological study, we simply incorporate a negative CC
into the parameterized $w_0w_a$CDM model, however, in realistic
scalar field model it might be that the corresponding DE has an
AdS vacuum, thus at low redshift it will roll towards the AdS
vacuum, so that our universe will collapse at certain redshift
$z<0$. However, relevant scalar field phenomenology seems not so
simple and might be richer than expected. The test for negative CC
with the high-redshift JWST massive galaxies has been investigated
in Refs.\cite{Adil:2023ara,Menci:2024rbq}. How to combine the JWST
and Planck+DESI+Pantheon+ data to search for the negative CC is
also an interesting issue.

It should be mentioned that we also have considered our
Planck18+DESI+Pantheon+ dataset combined with the SH0ES prior,
which, however, only suggests $H_0\simeq 70$ \cite{Wang202409},
thus it seems that such a low-redshift negative CC is not enough
to solve the Hubble tension, and a pre-recombination resolution
might be still necessary, see also \cite{Giare:2024syw} for a
recent model-independent test. Thus it is worth re-accessing the
corresponding results in the Hubble-tension-free cosmologies, as
in recent Refs.\cite{Wang:2024dka,Seto:2024cgo}.


\section*{Acknowledgments}
This work is supported by NSFC, No.12075246, National Key Research
and Development Program of China, No. 2021YFC2203004, and the
Fundamental Research Funds for the Central Universities.

\appendix

\section{MCMC results of $w_0w_a$CDM+CC model}\label{appendix}

\begin{figure*}
    \includegraphics[width=1.6\columnwidth]{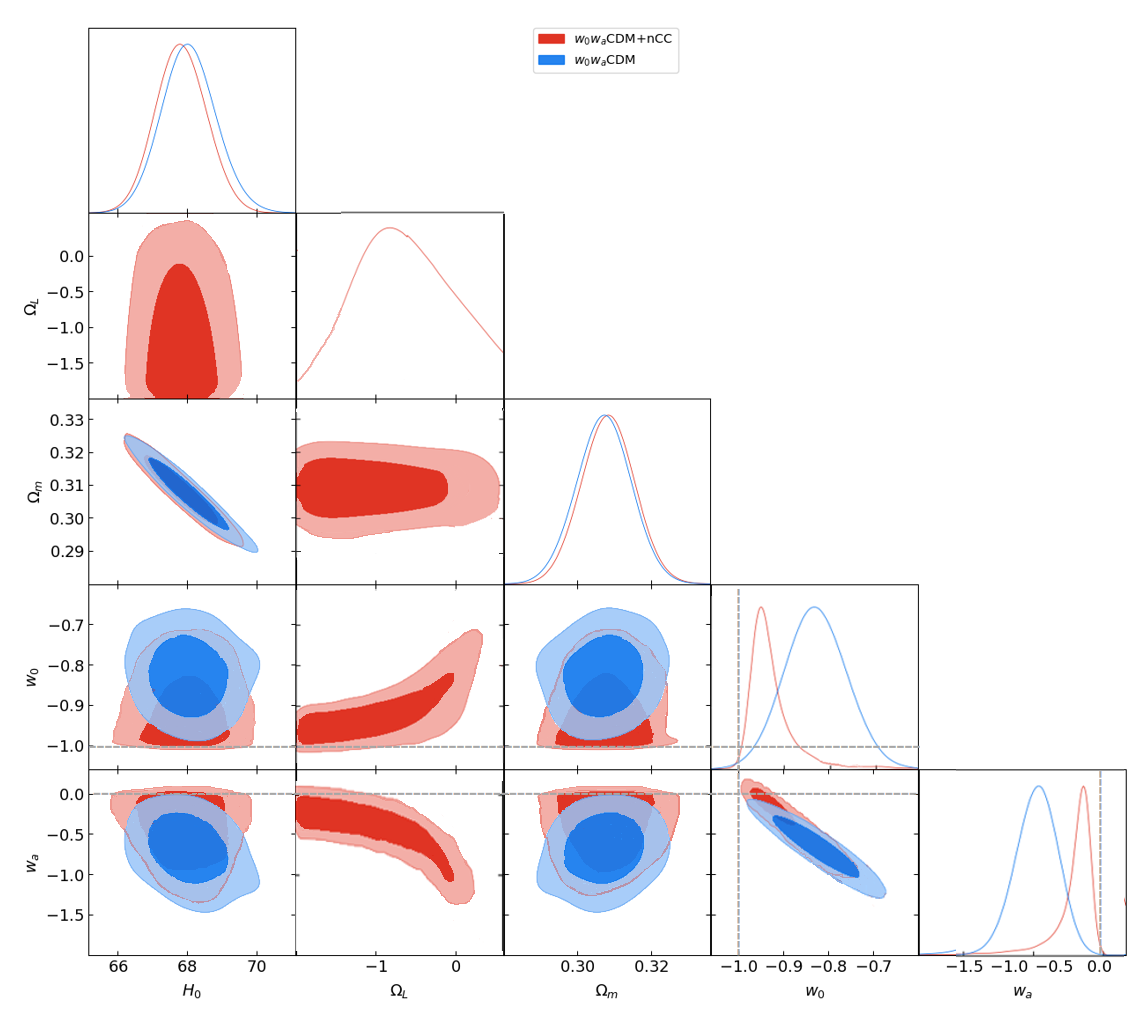}
    \caption{\label{MC}2D contours at 68\% and 95\% CL for the
parameters of the $w_0w_a$CDM+CC model. The dataset used is the
Planck18+DESI+Pantheon Plus dataset.}
\end{figure*}

In Fig.\ref{MC}, we present the $1\sigma$ and $2\sigma$
marginalized posterior distributions of main parameters for the
$w_0w_a$CDM+CC model.

\begin{thebibliography}{99}


\bibitem{Frieman:2008sn}
J.~Frieman, M.~Turner and D.~Huterer,
Ann. Rev. Astron. Astrophys. \textbf{46} (2008), 385-432
doi:10.1146/annurev.astro.46.060407.145243
[arXiv:0803.0982 [astro-ph]].

\bibitem{Nojiri:2017ncd}
S.~Nojiri, S.~D.~Odintsov and V.~K.~Oikonomou,
Phys. Rept. \textbf{692} (2017), 1-104
doi:10.1016/j.physrep.2017.06.001
[arXiv:1705.11098 [gr-qc]].

\bibitem{Carroll:2000fy}
S.~M.~Carroll,
Living Rev. Rel. \textbf{4} (2001), 1
doi:10.12942/lrr-2001-1
[arXiv:astro-ph/0004075 [astro-ph]].

\bibitem{Huterer:2017buf}
D.~Huterer and D.~L.~Shafer,
Rept. Prog. Phys. \textbf{81} (2018) no.1, 016901
doi:10.1088/1361-6633/aa997e
[arXiv:1709.01091 [astro-ph.CO]].

\bibitem{Weinberg:1988cp}
S.~Weinberg,
Rev. Mod. Phys. \textbf{61} (1989), 1-23
doi:10.1103/RevModPhys.61.1

\bibitem{Padmanabhan:2002ji}
T.~Padmanabhan,
Phys. Rept. \textbf{380} (2003), 235-320
doi:10.1016/S0370-1573(03)00120-0
[arXiv:hep-th/0212290 [hep-th]].


\bibitem{DESI:2024uvr}
A.~G.~Adame \textit{et al.} [DESI],
[arXiv:2404.03000 [astro-ph.CO]].

\bibitem{DESI:2024lzq}
A.~G.~Adame \textit{et al.} [DESI],
[arXiv:2404.03001 [astro-ph.CO]].

\bibitem{DESI:2024mwx}
A.~G.~Adame \textit{et al.} [DESI],
[arXiv:2404.03002 [astro-ph.CO]].

\bibitem{Chevallier:2000qy}
M.~Chevallier and D.~Polarski,
Int. J. Mod. Phys. D \textbf{10} (2001), 213-224
doi:10.1142/S0218271801000822 [arXiv:gr-qc/0009008 [gr-qc]].

\bibitem{Linder:2002et}
E.~V.~Linder,
Phys. Rev. Lett. \textbf{90} (2003), 091301
doi:10.1103/PhysRevLett.90.091301 [arXiv:astro-ph/0208512
[astro-ph]].

\bibitem{DESI:2024aqx}
R.~Calderon \textit{et al.} [DESI],
[arXiv:2405.04216 [astro-ph.CO]].

\bibitem{DESI:2024kob}
K.~Lodha \textit{et al.} [DESI],
[arXiv:2405.13588 [astro-ph.CO]].

\bibitem{Yang:2024kdo}
Y.~Yang, X.~Ren, Q.~Wang, Z.~Lu, D.~Zhang, Y.~F.~Cai and
E.~N.~Saridakis,
[arXiv:2404.19437 [astro-ph.CO]].

\bibitem{Wang:2024dka}
H.~Wang and Y.~S.~Piao,
[arXiv:2404.18579 [astro-ph.CO]].

\bibitem{Wang:2024pui}
Z.~Wang, S.~Lin, Z.~Ding and B.~Hu,
[arXiv:2405.02168 [astro-ph.CO]].



\bibitem{Yin:2024hba}
W.~Yin,
[arXiv:2404.06444 [hep-ph]].

\bibitem{Luongo:2024fww}
O.~Luongo and M.~Muccino,
[arXiv:2404.07070 [astro-ph.CO]].

\bibitem{Cortes:2024lgw}
M.~Cort\^es and A.~R.~Liddle,
[arXiv:2404.08056 [astro-ph.CO]].


        \bibitem{Carloni:2024zpl}
        Y.~Carloni, O.~Luongo and M.~Muccino,
        [arXiv:2404.12068 [astro-ph.CO]].

\bibitem{Wang:2024hks}
D.~Wang,
[arXiv:2404.06796 [astro-ph.CO]].

        \bibitem{Wang:2024rjd}
        D.~Wang,
        [arXiv:2404.13833 [astro-ph.CO]].

        \bibitem{Colgain:2024xqj}
        E.~\'O.~Colg\'ain, M.~G.~Dainotti, S.~Capozziello, S.~Pourojaghi, M.~M.~Sheikh-Jabbari and D.~Stojkovic,
        [arXiv:2404.08633 [astro-ph.CO]].

\bibitem{Giare:2024smz}
W.~Giar\`e, M.~A.~Sabogal, R.~C.~Nunes and E.~Di Valentino,
[arXiv:2404.15232 [astro-ph.CO]].

\bibitem{Gomez-Valent:2024tdb}
A.~Gomez-Valent and J.~Sola Peracaula,
[arXiv:2404.18845 [astro-ph.CO]].

\bibitem{Escamilla-Rivera:2024sae}
C.~Escamilla-Rivera and R.~Sandoval-Orozco,
JHEAp \textbf{42} (2024), 217-221 doi:10.1016/j.jheap.2024.05.005
[arXiv:2405.00608 [astro-ph.CO]].

\bibitem{Park:2024jns}
C.~G.~Park, J.~de Cruz Perez and B.~Ratra,
[arXiv:2405.00502 [astro-ph.CO]].

\bibitem{Shlivko:2024llw}
D.~Shlivko and P.~Steinhardt,
[arXiv:2405.03933 [astro-ph.CO]].


\bibitem{Dinda:2024kjf}
B.~R.~Dinda,
[arXiv:2405.06618 [astro-ph.CO]].

\bibitem{Seto:2024cgo}
O.~Seto and Y.~Toda,
[arXiv:2405.11869 [astro-ph.CO]].

\bibitem{Bhattacharya:2024hep}
S.~Bhattacharya, G.~Borghetto, A.~Malhotra, S.~Parameswaran,
G.~Tasinato and I.~Zavala,
[arXiv:2405.17396 [astro-ph.CO]].

\bibitem{Wolf:2023uno}
W.~J.~Wolf and P.~G.~Ferreira,
Phys. Rev. D \textbf{108} (2023) no.10, 103519
doi:10.1103/PhysRevD.108.103519 [arXiv:2310.07482 [astro-ph.CO]].

\bibitem{Kachru:2003aw}
S.~Kachru, R.~Kallosh, A.~D.~Linde and S.~P.~Trivedi,
Phys. Rev. D \textbf{68} (2003), 046005
doi:10.1103/PhysRevD.68.046005 [arXiv:hep-th/0301240 [hep-th]].

\bibitem{Kallosh:2004yh}
R.~Kallosh and A.~D.~Linde,
JHEP \textbf{12} (2004), 004 doi:10.1088/1126-6708/2004/12/004
[arXiv:hep-th/0411011 [hep-th]].

\bibitem{Ooguri:2006in}
H.~Ooguri and C.~Vafa,
Nucl. Phys. B \textbf{766} (2007), 21-33
doi:10.1016/j.nuclphysb.2006.10.033 [arXiv:hep-th/0605264
[hep-th]].

\bibitem{Obied:2018sgi}
G.~Obied, H.~Ooguri, L.~Spodyneiko and C.~Vafa,
[arXiv:1806.08362 [hep-th]].

\bibitem{Palti:2019pca}
E.~Palti,
Fortsch. Phys. \textbf{67} (2019) no.6, 1900037
doi:10.1002/prop.201900037
[arXiv:1903.06239 [hep-th]].

\bibitem{Grana:2021zvf}
M.~Gra\~na and A.~Herr\'aez,
Universe \textbf{7} (2021) no.8, 273
doi:10.3390/universe7080273
[arXiv:2107.00087 [hep-th]].

\bibitem{Kallosh:2019axr}
R.~Kallosh, A.~Linde, E.~McDonough and M.~Scalisi,
JHEP \textbf{03} (2019), 134 doi:10.1007/JHEP03(2019)134
[arXiv:1901.02022 [hep-th]].



\bibitem{Ye:2020btb}
G.~Ye and Y.~S.~Piao,
Phys. Rev. D \textbf{101} (2020) no.8, 083507
doi:10.1103/PhysRevD.101.083507 [arXiv:2001.02451 [astro-ph.CO]].


\bibitem{Ye:2020oix}
G.~Ye and Y.~S.~Piao,
Phys. Rev. D \textbf{102} (2020) no.8, 083523
doi:10.1103/PhysRevD.102.083523 [arXiv:2008.10832 [astro-ph.CO]].

\bibitem{Jiang:2021bab}
J.~Q.~Jiang and Y.~S.~Piao,
Phys. Rev. D \textbf{104} (2021) no.10, 103524
doi:10.1103/PhysRevD.104.103524 [arXiv:2107.07128 [astro-ph.CO]].

\bibitem{Ye:2021iwa}
G.~Ye, J.~Zhang and Y.~S.~Piao,
Phys. Lett. B \textbf{839} (2023), 137770
doi:10.1016/j.physletb.2023.137770 [arXiv:2107.13391
[astro-ph.CO]].

\bibitem{Wang:2022jpo}
H.~Wang and Y.~S.~Piao,
Phys. Lett. B \textbf{832} (2022), 137244
doi:10.1016/j.physletb.2022.137244 [arXiv:2201.07079
[astro-ph.CO]].

\bibitem{Akarsu:2019hmw}
\"O.~Akarsu, J.~D.~Barrow, L.~A.~Escamilla and J.~A.~Vazquez,
Phys. Rev. D \textbf{101} (2020) no.6, 063528
doi:10.1103/PhysRevD.101.063528 [arXiv:1912.08751 [astro-ph.CO]].

\bibitem{Akarsu:2021fol}
\"O.~Akarsu, S.~Kumar, E.~\"Oz\"ulker and J.~A.~Vazquez,
Phys. Rev. D \textbf{104} (2021) no.12, 123512
doi:10.1103/PhysRevD.104.123512 [arXiv:2108.09239 [astro-ph.CO]].

\bibitem{Akarsu:2022typ}
O.~Akarsu, S.~Kumar, E.~\"Oz\"ulker, J.~A.~Vazquez and A.~Yadav,
Phys. Rev. D \textbf{108} (2023) no.2, 023513
doi:10.1103/PhysRevD.108.023513 [arXiv:2211.05742 [astro-ph.CO]].

\bibitem{Akarsu:2023mfb}
O.~Akarsu, E.~Di Valentino, S.~Kumar, R.~C.~Nunes, J.~A.~Vazquez
and A.~Yadav,
[arXiv:2307.10899 [astro-ph.CO]].


\bibitem{Paraskevas:2024ytz}
E.~A.~Paraskevas, A.~Cam, L.~Perivolaropoulos and O.~Akarsu,
Phys. Rev. D \textbf{109} (2024) no.10, 103522
doi:10.1103/PhysRevD.109.103522 [arXiv:2402.05908 [astro-ph.CO]].

\bibitem{Anchordoqui:2023woo}
L.~A.~Anchordoqui, I.~Antoniadis and D.~Lust,
[arXiv:2312.12352 [hep-th]].

\bibitem{Li:2019ipk}
H.~H.~Li, G.~Ye, Y.~Cai and Y.~S.~Piao,
Phys. Rev. D \textbf{101} (2020) no.6, 063527
doi:10.1103/PhysRevD.101.063527 [arXiv:1911.06148 [gr-qc]].

\bibitem{Ye:2022efx}
G.~Ye, J.~Q.~Jiang and Y.~S.~Piao,
Phys. Rev. D \textbf{106} (2022) no.10, 103528
doi:10.1103/PhysRevD.106.103528 [arXiv:2205.02478 [astro-ph.CO]].

\bibitem{Lin:2022ygd}
P.~X.~Lin, H.~L.~Huang, J.~Zhang and Y.~S.~Piao,
to be published in Phys.Lett.B [arXiv:2211.05265 [gr-qc]].

\bibitem{Piao:2005ag}
Y.~S.~Piao,
Phys. Rev. D \textbf{71} (2005), 087301
doi:10.1103/PhysRevD.71.087301 [arXiv:astro-ph/0502343
[astro-ph]].

\bibitem{Cai:2017pga}
Y.~Cai, Y.~T.~Wang, J.~Y.~Zhao and Y.~S.~Piao,
Phys. Rev. D \textbf{97} (2018) no.10, 103535
doi:10.1103/PhysRevD.97.103535 [arXiv:1709.07464 [astro-ph.CO]].



\bibitem{Dutta:2018vmq}
K.~Dutta, Ruchika, A.~Roy, A.~A.~Sen and M.~M.~Sheikh-Jabbari,
Gen. Rel. Grav. \textbf{52} (2020) no.2, 15
doi:10.1007/s10714-020-2665-4
[arXiv:1808.06623 [astro-ph.CO]].

\bibitem{Visinelli:2019qqu}
L.~Visinelli, S.~Vagnozzi and U.~Danielsson,
Symmetry \textbf{11} (2019) no.8, 1035
doi:10.3390/sym11081035
[arXiv:1907.07953 [astro-ph.CO]].

\bibitem{Ruchika:2020avj}
Ruchika, S.~A.~Adil, K.~Dutta, A.~Mukherjee and A.~A.~Sen,
Phys. Dark Univ. \textbf{40} (2023), 101199
doi:10.1016/j.dark.2023.101199
[arXiv:2005.08813 [astro-ph.CO]].


\bibitem{Calderon:2020hoc}
R.~Calder\'on, R.~Gannouji, B.~L'Huillier and D.~Polarski,
Phys. Rev. D \textbf{103} (2021) no.2, 023526
doi:10.1103/PhysRevD.103.023526
[arXiv:2008.10237 [astro-ph.CO]].

\bibitem{Sen:2021wld}
A.~A.~Sen, S.~A.~Adil and S.~Sen,
Mon. Not. Roy. Astron. Soc. \textbf{518} (2022) no.1, 1098-1105
doi:10.1093/mnras/stac2796
[arXiv:2112.10641 [astro-ph.CO]].

\bibitem{Malekjani:2023ple}
M.~Malekjani, R.~M.~Conville, E.~\'O.~Colg\'ain, S.~Pourojaghi and M.~M.~Sheikh-Jabbari,
Eur. Phys. J. C \textbf{84} (2024) no.3, 317
doi:10.1140/epjc/s10052-024-12667-z
[arXiv:2301.12725 [astro-ph.CO]].

\bibitem{Gomez-Valent:2023uof}
A.~G\'omez-Valent, A.~Favale, M.~Migliaccio and A.~A.~Sen,
Phys. Rev. D \textbf{109} (2024) no.2, 023525
doi:10.1103/PhysRevD.109.023525 [arXiv:2309.07795 [astro-ph.CO]].

\bibitem{Adil:2023ara}
S.~A.~Adil, U.~Mukhopadhyay, A.~A.~Sen and S.~Vagnozzi,
JCAP \textbf{10} (2023), 072
doi:10.1088/1475-7516/2023/10/072
[arXiv:2307.12763 [astro-ph.CO]].

\bibitem{Menci:2024rbq}
N.~Menci, S.~A.~Adil, U.~Mukhopadhyay, A.~A.~Sen and S.~Vagnozzi,
[arXiv:2401.12659 [astro-ph.CO]].




\bibitem{Planck:2018vyg}
N.~Aghanim \textit{et al.} [Planck],
Astron. Astrophys. \textbf{641}, A6 (2020)
[erratum: Astron. Astrophys. \textbf{652}, C4 (2021)]
doi:10.1051/0004-6361/201833910
[arXiv:1807.06209 [astro-ph.CO]].

\bibitem{Planck:2019nip}
N.~Aghanim \textit{et al.} [Planck],
Astron. Astrophys. \textbf{641} (2020), A5
doi:10.1051/0004-6361/201936386
[arXiv:1907.12875 [astro-ph.CO]].

\bibitem{Planck:2018lbu}
N.~Aghanim \textit{et al.} [Planck],
Astron. Astrophys. \textbf{641} (2020), A8
doi:10.1051/0004-6361/201833886
[arXiv:1807.06210 [astro-ph.CO]].


\bibitem{Scolnic:2021amr}
D.~Scolnic, D.~Brout, A.~Carr, A.~G.~Riess, T.~M.~Davis, A.~Dwomoh, D.~O.~Jones, N.~Ali, P.~Charvu and R.~Chen, \textit{et al.}
Astrophys. J. \textbf{938} (2022) no.2, 113
doi:10.3847/1538-4357/ac8b7a
[arXiv:2112.03863 [astro-ph.CO]].

\bibitem{Audren:2012wb}
B.~Audren, J.~Lesgourgues, K.~Benabed and S.~Prunet,
JCAP \textbf{02} (2013), 001
doi:10.1088/1475-7516/2013/02/001
[arXiv:1210.7183 [astro-ph.CO]].

\bibitem{Brinckmann:2018cvx}
T.~Brinckmann and J.~Lesgourgues,
Phys. Dark Univ. \textbf{24} (2019), 100260
doi:10.1016/j.dark.2018.100260
[arXiv:1804.07261 [astro-ph.CO]].

\bibitem{Lesgourgues:2011re}
J.~Lesgourgues,
[arXiv:1104.2932 [astro-ph.IM]].



\bibitem{Blas:2011rf}
D.~Blas, J.~Lesgourgues and T.~Tram,
JCAP \textbf{07} (2011), 034
doi:10.1088/1475-7516/2011/07/034
[arXiv:1104.2933 [astro-ph.CO]].




\bibitem{Ozulker:2022slu}
E.~Ozulker,
Phys. Rev. D \textbf{106} (2022) no.6, 063509
doi:10.1103/PhysRevD.106.063509 [arXiv:2203.04167 [astro-ph.CO]].

\bibitem{Wang202409}
H. Wang et.al to be submitted.

\bibitem{Giare:2024syw}
W.~Giar\`e, J.~Betts, C.~van de Bruck and E.~Di Valentino,
[arXiv:2406.07493 [astro-ph.CO]].








\end{thebibliography}
\end{document}